# SUPPRESSION OF SCATTERING FOR SMALL DIELECTRIC PARTICLES: AN ANAPOLE MODE AND INVISIBILITY

Boris Luk`yanchuk[1*], Ramón Paniagua-Domínguez[1], Arseniy I. Kuznetsov[1],
Andrey E. Miroshnichenko[2] and Yuri S. Kivshar[2]

[1]*Data Storage Institute, A*STAR (Agency for Science, Technology and Research), 2 Fusionopolis Way, #08-01, Innovis 138634, Singapore.*
[2]*Nonlinear Physics Centre, Research School of Physics and Engineering, Australian National University, Canberra ACT 2601, Australia.*

**We reveal that an isotropic homogeneous subwavelength particle with a high refractive index can produce ultra-weak total scattering due to vanishing contribution of the electric dipole moment. This effect can be explained with the help of the Fano resonance and scattering efficiency associated with the excitation of an anapole mode. The latter is a nonradiative mode emerging from destructive interference of electric and toroidal dipole moments, and it can be employed for a design of highly transparent optical materials.**

## 1. Introduction

The idea of perfect invisibility follows the history of humanity. This idea inspired poets and writers who described great advantages of invisibility for military, political, or even for love affairs. We may mention the numerous stories starting from the Arabian "*One Thousand and One Nights*", Alexander Pushkin`s poem "*Ruslan and Ludmila*", science fiction novels such as "*The Invisible Man*" by Herbert G. Wells, and more recent stories such as *Harry Potter* of Joanne Rowling and many others. However, for a long time this invisibility phenomenon was considered to be forbidden by laws of general physics. Since the time of Lord Rayleigh[1], it was recognized that even a very small particle will be visible due to light scattering. Attempts in Nature, e.g. for chameleons, to realize invisibility based on camouflage (i.e. imitating colors of a background) and, thus, are just a palliative.

In a book published in 1962, P. Ufimtsev suggested another principle based on the use of a special shape of an aircraft to escape back reflection of radar signals. Such a property is demonstrated, for example, with a conical mirror directed toward a light source. Ufimtsev`s book was translated to English[2], and it gave a start for the development of the so-called stealth technology. It is interesting to mention that M. Levin (who was an external examiner of Ufimtsev`s PhD thesis) wrote in his official report that the zero backscattering could be derived for an azimuthally symmetric material with $\varepsilon = \mu$ (Ufimtsev highlighted this story in a book[3]). Meanwhile, the same idea was later rediscovered[4] and published by M. Kerker, after which this condition is named as the first Kerker condition, confirming the general principle "*if a notion bears a personal name, then this name is not the name of the discoverer*"[5]. However, Kerker developed further the idea how to suppress the total scattering. For that, he considered the scattering from multi-layered spheres[6] and spheroids[7].

A step further in the development of the invisibility concept led to the next advance based on the concept of cloaking[8,9] which uses general principles of transformation optics such as the Fermat principle. This idea is based on the creation of inhomogeneous media that forces light to go around the cloaking area. Consequently, it is necessary to create some special media with varying permittivity $\varepsilon = \varepsilon(\mathbf{r})$ and permeability $\mu = \mu(\mathbf{r})$ functions, which make practical realizations rather difficult. Other ideas related to multi-shell structures for cloaking have been developed for plasmonic[10], dielectric[11] and metamaterial coatings together with scattering cancellation and mantle cloaking[12].

Returning to homogeneous spheres, we should mention the idea of directional scattering for plasmonic nanoparticles closely related to the Fano resonance in plasmonic materials and metamaterials[13, 14]. However, such Fano resonances and directional suppression of scattering in plasmonic particles is not accompanied by minimization of the total scattering efficiency. Another problem is related to dissipation of real metals, which can be easily circumvented if dielectric particles are used instead.

## 2. Rayleigh approximation

Scattering efficiency, $Q_{sca}$, in the Rayleigh scattering regime is given by the well-known formula[15]

$$Q_{sca}^{(Ra)} = \frac{8}{3}\left(\frac{\varepsilon-1}{\varepsilon+2}\right)^2 q^4, \qquad (1)$$

representing a ratio of the scattering cross-section to the geometrical cross-section, $\sigma_{geom} = \pi R^2$, where $R$ is particle radius, $\varepsilon$ is its permittivity, and $q = 2\pi R/\lambda$ is the so called size parameter, with $\lambda$ the radiation wavelength. The particle is considered to be a small ideal sphere made of an isotropic, homogeneous and nonmagnetic ($\mu = 1$) material. In addition, the size of this particle should be sufficiently small. If one consider non-dissipative, $\text{Im}\,\varepsilon = 0$, dielectric materials with positive refractive index $n = \sqrt{\varepsilon} > 1$, then the ratio of scattering efficiency to the fourth power of the size parameter presents a universal function, which monotonously increases with the refractive index

$$Q_{sca}^{(Ra)}/q^4 = \frac{8}{3}\left(\frac{n^2-1}{n^2+2}\right)^2. \qquad (2)$$

Formula (1) represents the scattering of electrical dipole. Justification of this formula follows from the Mie theory[15], which is the exact solution of Maxwell's equations for scattering of a plane wave from a spherical particle. Scattering efficiency of the particle according to Mie theory is given by

$$Q_{sca} = \frac{2}{q^2}\sum_{\ell=1}^{\infty}(2\ell+1)\left[|a_\ell|^2 + |b_\ell|^2\right], \qquad (3)$$

where the scattering amplitudes $a_\ell$ (electric) and $b_\ell$ (magnetic) are expressed in terms of the Ricatti-Bessel functions[15]. With small size parameter, $q \ll 1$, one can find[16] $a_\ell \propto q^{2\ell+1}$ and $b_\ell \propto q^{2\ell+3}$. Thus, the electrical dipole amplitude $a_1$ is dominant. For a small plasmonic particle with $\varepsilon < 0$ the scattering efficiency can be very large, $Q_{sca} \gg 1$. The physical reason for this effect can be explained from the Poynting vector field[16, 17], which indicates that the cross-section of separatrix tubes for the energy flow into the particle can greatly exceed the geometrical cross-section. Near plasmonic resonances with $\varepsilon = -(1+\ell^{-1})$, Rayleigh approximation is not valid, e.g. $Q_{sca}^{(Ra)}$ in formula (1) has singularity at $\varepsilon = -2$. However, the Mie theory yields for electrical dipole resonance a limited value $Q_{sca} = 6/q^2$. For weakly dissipative plasmonic materials, one can see an inversion of the hierarchy of resonances, when the scattering efficiency at the dipole resonance is smaller than that of the quadrupole one, which is, in turn, smaller than for the octupole resonance, etc.[18]

A small dielectric particle with $\varepsilon > 1$ produces, according to Rayleigh approximation (1), a very small scattering, which tends to zero at $q \to 0$. This scattering corresponds to the situation when the electrical dipole amplitude plays the dominant role and all other amplitudes are small. Such situation is typical for small plasmonic particles. A similar situation takes place for small size parameter $q \ll 1$ and refractive index $1 < n < 2$, see e.g.[19]. For these cases, Rayleigh formula (1) represents the minimal possible scattering of a small particle.

## 3. High-index dielectric particles

This situation changes for a particle with high refractive index. For example, silicon particles at the optical range $(n \approx 4)$ have scattering efficiencies at the magnetic dipole resonance that are larger than at the electrical dipole resonance even for small particles[20,21], as has been experimentally confirmed[22, 23]. As a result, the scattering near the resonances is not small in spite of fulfilling the condition that the particle size is small. In fact, the true conditions for the applicability of formula (1) are $q \ll 1$ and, additionally, $q < 1/n$. This is easy to see in Fig. 1a, where the total Mie scattering and Rayleigh scattering efficiencies versus refractive index $n$ are presented for a spherical particle with size parameter $q = 0.3$ $(R/\lambda \approx 0.05)$. Rayleigh scattering saturates at large $n$, but formula (1) loses its validity in the vicinity of the magnetic dipole resonance at $n = 10.3$ and also in the vicinity of the subsequent resonances (as shown in the inset of Fig.1a). Note that some very narrow peaks cannot be seen on the scale of this inset. Also note that, even with large refractive index, the total scattering between resonances is quite close to Rayleigh scattering (blue line).

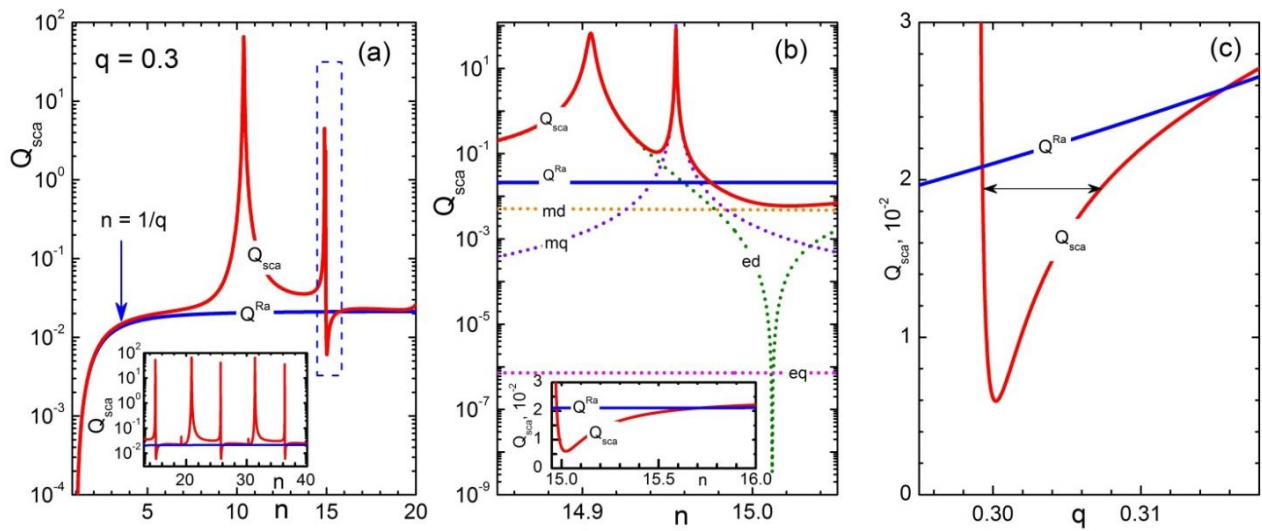

Fig. 1. (a) Scattering efficiencies according to Rayleigh approximation (blue line) and exact Mie theory (red line) for small size parameter $q = 0.3$. The inset in (a) shows resonances at higher values of the refractive index. In the vicinity of $n = 15, 21, 36…$ one can see that the total scattering is less than it follows from formula (1). A zoom of the scattering within the area indicated by the dashed box is shown in (b), where Rayleigh approximation (blue line) and exact Mie theory (red line) are shown together with four partial scattering efficiencies for magnetic dipole (md), electric dipole (ed), magnetic quadrupole (md) and electric quadrupole (eq). Insets in (b) and (c) show $Q_{sca} < Q^{Ra}$ region in linear coordinates.

It is not surprising that the particle has a scattering efficiency much larger than Rayleigh scattering near resonances. However, the asymmetric line-shape of the resonance (see Fig 1a) leads also to a strong suppression of total scattering near the resonance that becomes lower than that given by Rayleigh formula (1). Such scattering behaviour is achieved e.g. in the region near $n \approx 15$ marked in bracket in Fig. 1a. The zoom into this area is shown in Fig. 1b. There are two closely situated electric dipole and magnetic quadrupole resonances, which cannot be resolved on the scale of Fig. 1a. Although there are many points with local scattering minima, as shown in the inset to Fig. 1a, a more precise examination shows that the global minimum in scattering is reached near $n \approx 15$. Detailed analysis of this resonance reveals that it has a typical Fano shape. Typically, Fano bandwidth is quite narrow, see in Fig. 1c.

Previously, it was shown that such Fano shape arises in the total scattering from single elongated antennas, both in the plasmonic case[24, 25] and in their dielectric analoques[26], within a system of disordered photonic crystals[27] as well as in the transmission spectra of two-dimensional square lattices of dielectric circular rods[28]. In the last case, the Fano resonance is arising due to interplay between the resonant Mie scattering from individual rods and the Bragg scattering from the photonic lattice. One should also mention the Fano profile of the dipole scattering amplitude $|a_1(n)|^2$ at the limit $n \gg 1$[29]. Formally, results of Ref. 29 can be expressed as an interference of two partitions, where one corresponds to the $n$-independent wave, scattered by a perfectly reflecting particle and plays the role of a background, while the other is associated with the excitation of a resonant Mie mode[30].

The Fano resonance shape in our case is associated with the destructive interference of an electric dipole with a toroidal dipole mode. Previously it was shown that such type of interference, the so-called anapole mode[31, 32], could be observed in Si nanodisks. For nanodisks, this anapole mode can be achieved at a specific wavelength and for some fixed ratio of the disk height to its diameter. It was shown as well that, in the case of a single isolated nonmagnetic isotropic spherical particle, the anapole condition is usually hidden by the rest of multipolar contributions, difficulting its observation. Here, we show that the anapole excitation may be observed even in the simple spherical case, provided the particle is sufficiently small and has a sufficiently high refractive index. In Fig. 2 we present the Poynting vector distribution for the particle with $n = 15.0116$ and $q = 0.3$. The total scattering efficiency, $Q_{sca} \approx 5.99 \cdot 10^{-3}$, for this particle is about 3.5 times less than Rayleigh scattering calculated by formula (1), $Q_{sca}^{(Ra)} \approx 2.11 \cdot 10^{-2}$. One can see in Fig. 2 that the Poynting vector has a toroidal structure. Note that at $q \ll 1$ the magnetic dipole contribution is small, so that inequality $Q_1^{(m)} \ll Q_1^{(e)}$ is satisfied at almost all values of parameters except of regions close to the anapole conditions. Closed singular line in Fig. 2b corresponds to zero energy flow. Near this singular line Poynting vector produces characteristic vortices, see in Fig. 2c, similar to vortices in small plasmonic particle[16, 33].

Additionally we can prove the toroidal symmetry by plotting distributions of electric and magnetic vectors inside the particle in mutually perpendicular planes as it is shown in Fig. 3. Here distribution of electric vector **E** is shown within the $\{x,z\}$ plane through the diameter of the particle in Fig. 3a. Colour panel in Fig. 3a presents the intensity $\mathbf{E}^2$ distribution. Within the perpendicular $\{y,z\}$ plane, one can see the distribution of magnetic vector **H**, as shown in Fig. 3b. Similar toroidal fields[32] recently attract a lot of attention. Dominant contributions of electric and toroidal dipole moments can be clearly seen in Cartesian coordinates[31]. Thus, one can conclude that this Fano resonance is related to constructive and destructive interference of electrical dipole and toroidal dipole moments.

As it follows from Fig. 1b the corresponding Fano resonance arises in the vicinity of the zero of electrical dipole mode, $a_1 = 0$. This condition yields the equation

$$1 - n^2 + q(n^2 - 1 + n^2 q^2)\cot(q) + nq(n^2 - 1 - n^2 q^2)\cot(nq) + nq^2(1 - n^2)\cot(q)\cot(nq) = 0. \qquad (4)$$

Within the equation (4) one should consider $\cos(q) \neq 0$ and $\cos(nq) \neq 0$. For each value of refractive index, there are infinite set of solutions with corresponding size parameter $q$. However just the first root corresponds to the global minimum of $Q_{sca}$.

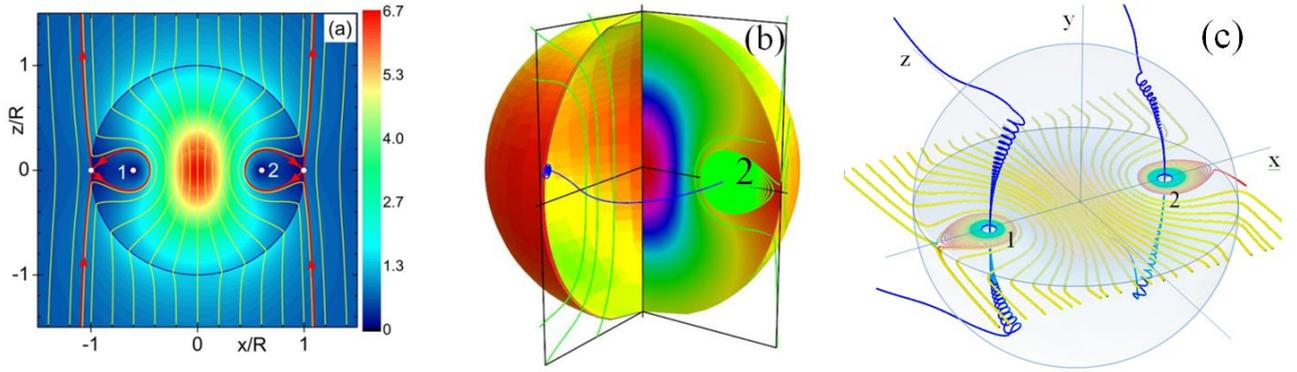

Fig. 2. (a) 2D Poynting vector in $\{x,z\}$ plane for $q=0.3$ and $n=15.0116$. Colour panel presents the variation of the modulus of the Poynting vector. The full number of modes in the Mie theory is taken into account. There are 2 saddles and 2 focal points. Red lines show the separatrices. Inside the particle, one can see the loops of separatrices representing the cross-sections of the electric toroidal dipole. (b) 3D Poynting vector distribution. The focal points in Fig. 2 are in reality unstable saddle-focal points. Through these points goes the closed singular line, which provides the axis of the toroidal mode. One fourth part of this line is shown by blue line. Green lines show the untwisted spiral of the Poynting vector in $\{x,z\}$ plane. (c) Vortices around the closed singular line, which provides the axis for toroidal mode.

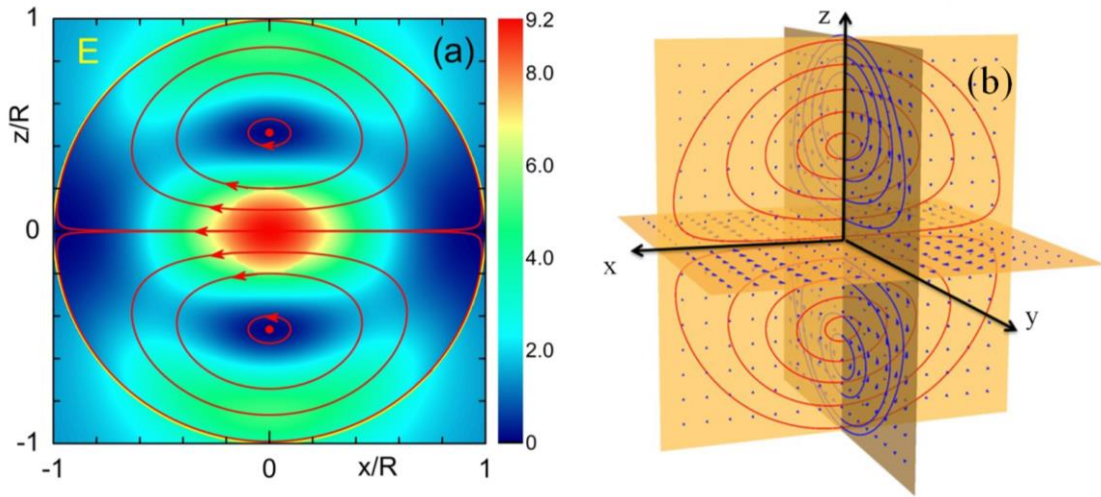

Fig. 3. (a) Distribution of electric vector **E** within the $\{x,z\}$ plane through the particle diameter. Colour panel indicates the value of electric intensity $\mathbf{E}^2$ in this plane. (b) 3D distributions of electric **E** (red lines) and magnetic **H** (blue lines) vectors within the planes through the particle diameter.

It is important to emphasize that the anapole mode corresponds to a global minimum in the scattering efficiency. Minimization of differential scattering functions[34-37] does not minimize the total scattering. For example, minimization of the forward scattering due to the second Kerker conditions[34] is quite close to local minimum in the scattering (see Fig. 4a), but it is still larger than the Rayleigh scattering. A better way to visualize the effect of the global minimization of scattering is shown in Fig. 4b, where the Rayleigh and anapole scattering efficiencies, $Q_{sca}/q^4$, are shown vs refractive index.

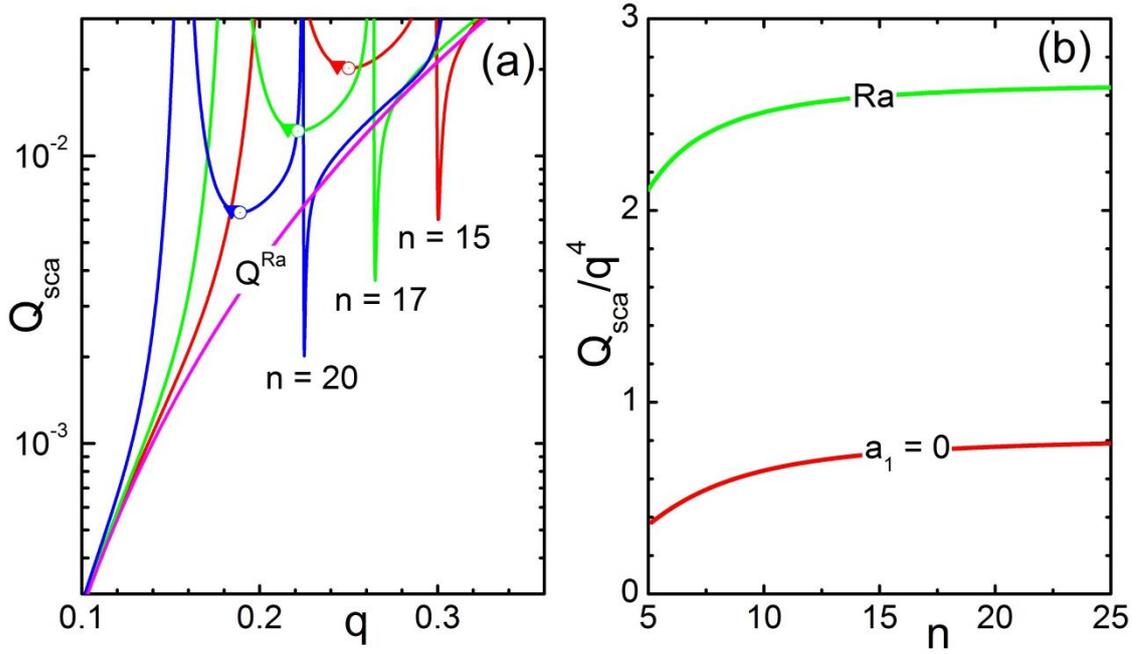

Fig. 4. (a) Scattering efficiencies vs. size parameter for: $n = 15$ (red), $n = 17$ (green), and $n = 20$ (blue). Open circles mark the positions of the local minima. Triangular marks show the scattering in the points which correspond to the minimal forward scattering at the second Kerker condition. (b) Function $Q_{sca}/q^4$ along the Rayleigh scattering (green) and anapole mode (red).

It is possible to produce further minimization of scattering using spheroidal particles. In contract to problem with scattering maximization[38] the answer for minimization in fact is quite evident and similar to the stealth effect: the needle directed toward to the light source will produce the minimal scattering.

It is also interesting to note that behaviour of directional scattering for spherical particle near the magnetic dipole resonance and anapole modes is quite different. The forward, $Q_{FS}$, and backward, $Q_{BS}$, scattering efficiencies defined for a spherical particle from Mie theory as:

$$Q_{FS} = \frac{1}{q^2}\left|\sum_{\ell=1}^{\infty}(2\ell+1)(a_\ell + b_\ell)\right|^2, \quad Q_{BS} = \frac{1}{q^2}\left|\sum_{\ell=1}^{\infty}(2\ell+1)(-1)^\ell(a_\ell - b_\ell)\right|^2. \quad (5)$$

A total suppression of the forward scattering is forbidden by the optical theorem[39]. The back scattering of small particle can be almost completely suppresses at the condition $a_1 = b_1$ which is referees as the *first Kerker condition*, while minimization of the inverse ratio is often referred to as the *second Kerker condition*[34]. Calculation with size parameter $q = 0.3$ yields the first Kerker condition for the particle with $n \approx 9.14$ and the second Kerker conditions for the particle with $n \approx 12.08$. It is interesting to note that the second branch of solutions for Kerker conditions at $q = 0.3$ yields the values $n \approx 14.99$ and $n \approx 15.14$ located in the vicinity of the anapole mode. The corresponding polar scattering diagrams[15] are shown in Fig. 5. At conventional Kerker conditions the scattering pattern is independent on incident polarization and is practically the same for linearly polarized and non-polarized light (i.e. it has rotational symmetry). In contrast, the scattering pattern near anapole is not rotationally symmetric, thus yielding polarization-dependent directional scattering (except at backward and forward directions). This behaviour is quite similar to the change of directivity in the vicinity of quadrupole resonance within the weakly dissipated plasmonic particles[14].

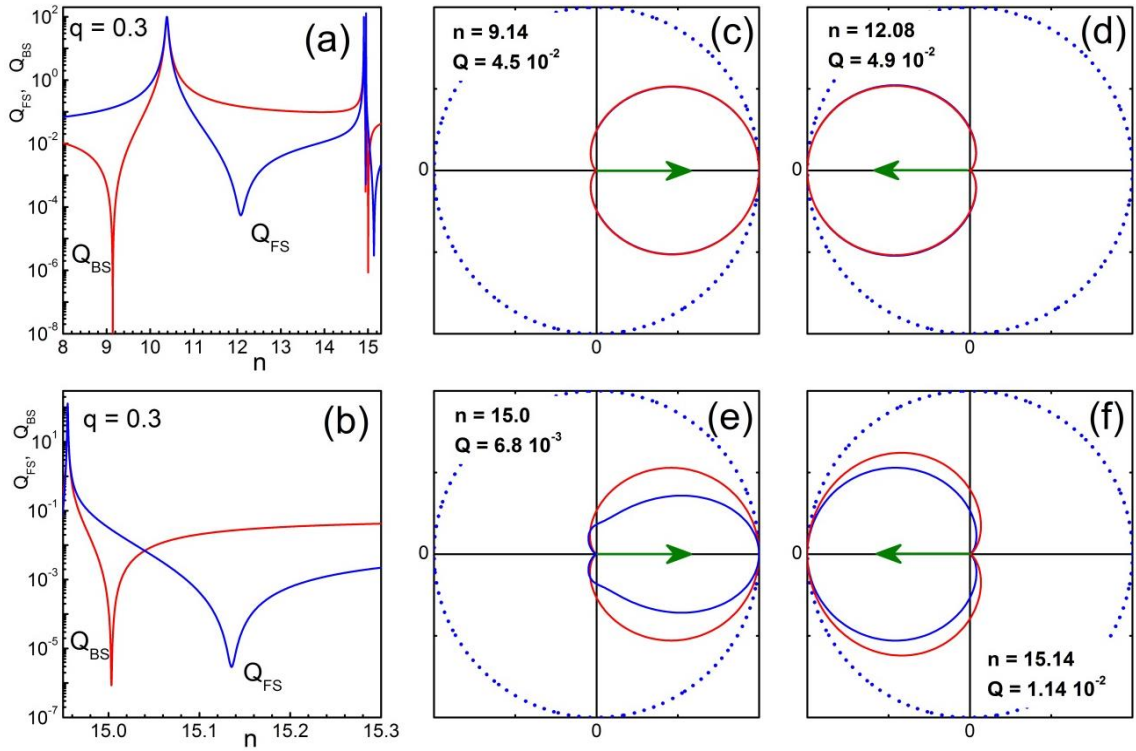

Fig. 5. Left panels (a, b) show forward and backward scattering efficiencies for small particle $q = 0.3$ versus refractive index $n$. The right panels (c-f) show the polar scattering diagrams in the x–z plane (azimuthal angle $\varphi = 0$ in Mie theory) for different refractive index $n$. Blue lines shows linearly polarized light and red lines represent non-polarized light. Arrows indicate direction of scattering.

## 4. Conclusions

We have found the conditions when the scattering efficiency of a small spherical dielectric particle is strictly below its value following from the Rayleigh formula (1). This effect of the scattering suppression is related to the excitation of an anapole mode associated with the Fano resonance. Similar effects were found previously for a homogeneous dielectric rod[30], but the physics behind was explained in a different way. It is important to highlight that our results refer to homogeneous isotropic small particles. The conditions for such ultra-weak scattering to occur are: a small value of the size parameter $q \ll 1$, large value of the refractive index typically $n > 5$, and weak dissipation. There are a number of materials satisfying these conditions in far-IR and RF frequency domains, e.g. SiC, $TiO_2$, ceramics, and other materials[40]. Also, clusters assembled of such nanoparticles may have interesting properties such as high transparency, i.e. the scattering effect in the extinction can be strongly suppressed.

## Acknowledgments

BL, RPD and AIK acknowledge support by DSI core funds and A*STAR Science and Engineering Research Council (SERC) Pharos grant #1527000025. AEM and YSK were supported by the Australian Research Council.

## Authors' Contributions

BL initiated the study of the anapole effects by small spherical particles. All authors contributed equally into discussions, writing, and editing this manuscript.